# A Hybrid Intrusion Detection with Decision Tree for Feature Selection

Mubarak Albarka Umar [1*], Chen Zhanfang [2], Yan Liu [3]

[1,2,3] School of Computer Science and Technology,

Changchun University of Science and Technology,

7186 Weixing Road, Jilin, China.

Corresponding Author: [*1]2018300037@mails.cust.edu.cn, [2]chenzhanfang@cust.edu.cn, [3]yanl@cust.edu.cn

**Abstract**

Due to the size and nature of intrusion detection datasets, intrusion detection systems (IDS) typically take high computational complexity to examine features of data and identify intrusive patterns. Data preprocessing techniques such as feature selection can be used to reduce such complexity by eliminating irrelevant and redundant features in the dataset. The objective of this study is to analyze the efficiency and effectiveness of some feature selection approaches namely, a wrapper-based and filter-based modeling approaches. To achieve that, a hybrid of feature selection algorithm in combination with wrapper and filter selection processes is designed. We propose a wrapper-based hybrid intrusion detection modeling with a decision tree algorithm to guide the selection process. Five machine learning algorithms are used on the wrapper and filter-based feature selection methods to build IDS models using the UNSW-NB15 dataset. The three filter-based methods namely, information gain, gain ratio, and relief are used for comparison to determine the efficiency and effectiveness of the proposed approach. Furthermore, a fair comparison with other state-of-the-art intrusion detection approaches is also performed. The experimental results show that our approach is quite effective in comparison to state-of-the-art works, however, it takes high computational time in comparison to the filter-based methods whilst achieves similar results. Our work also revealed unobserved issues about the conformity of the UNSW-NB15 dataset.

## 1  Introduction

Today more sophisticated infiltration techniques have been developed by attackers to challenge and defeat the security layer of the internet and computer users. Protecting the confidentiality, credibility, integrity, and availability of information communicate over the internet and across computers has become a vital and challenging task for network security administrators [1]. Thus, an efficient and reliable IDS is needed as an added security layer to the already existing less-effective first line of defense to safeguard computer networks from the known and unknown vulnerabilities [2], [3]. Machine Learning (ML) techniques, due to their ability to learn and improve with experience [4], are nowadays utilized for building such IDS [5]. There was one problem with the initial idea of applying ML in the form of a single classifier in IDS, that is, this approach may not be robust enough to build an effective IDS [6]. Thus, to enable building more accurate IDS, various researchers proposed hybrid IDS modeling approach to enhance the efficiency and effectiveness of detecting an intrusion [7].

An important aspect of building and validating IDS is the IDS dataset [8]. The dataset typically comes from heterogeneous platforms and can be noisy, redundant, incomplete, and inconsistent [9], this generally affects the detection accuracy and efficiency by increasing computational complexity and expanding the search space of the problem [10]. The primary purpose of IDS is to be accurate in detecting attacks with minimum false alerts. However, to fulfill this purpose, an IDS should be able to handle a huge amount of network data and should be fast enough to make real-time decisions. Preprocessing techniques such as normalization [11], [12], data filtration [6], discretization [13], and soon are used to overcome such issues. Feature selection is one of these techniques proposed by various researchers for increasing the detection efficiency and accuracy of IDS [7] and it has notably proven to be the most effective solution [14].

Feature selection is a process of selecting an optimal subset of features that infer the same meaning as the complete features set [14]. Feature selection aims to select relevant features and eliminates useless ones with a minimum or no degradation of performance. There are three main feature selection approaches, filter, wrapper, and embedded methods. The filter method extract features from data without using any learning algorithm, it relies on the general characteristics of the data such as distance, consistency, dependency, information, and correlation to evaluate and select feature subsets, it may however result in eliminating relevant and important features. The wrapper method uses a learning algorithm to determine the most useful and relevant features, and compared to the filter method, it improves model performance but it's computationally more expensive. Due to these limitations in each method, the embedded method was proposed to bridge the gap between the filter and wrapper methods. The embedded method performs feature selection during the learning time. In other words, it achieves model fitting and feature selection simultaneously [15].

In this work, which is an extension of our previous work [16], we propose a hybrid IDS approach for detecting network traffic intrusion. Wrapper-based feature selection with decision tree algorithm is first used as a regular means of obtaining an optimal subset of the original features, then five among the most used ML algorithms in IDS [5] are selected for building the models. The selected algorithms are Artificial Neural Network (ANN), k-Nearest Neighbor (KNN), Support Vector Machine (SVM), Random Forest (RF), and Naïve Bayes (NB). The models are implemented using the contemporary UNSW-NB15 dataset [8], [17] introduced by Moustafa and Slay [18]. One-hot encoding and min-max methods are used for encoding and normalization respectively. The three most used IDS eval metrics [5] namely, Accuracy, Detection rate, and False alert rate, are used in addition to the computation time, to evaluate the models. Furthermore, two comparisons are performed to determine the effectiveness of our method. Firstly, against three different feature selection methods namely, information gain (InfoGain), gain ratio (GainRatio), and Relief filters, and then against state-of-the-art works.

The rest of the paper is organized as follows. A literature review of the feature selection application in IDS is presented in Section 2. Section 3 introduces the general concept of feature selection; justification of the proposed feature selection is also given along with an explanation of the three other feature selection methods selected for comparison. Then, detailed experimental procedures using the proposed method as well as the filter methods is given in Section 4, while in Section 5 we provide the evaluation results, and discussions are made through comparative analysis on performance and computational time basis as well as with the state-of-the-art results obtained by researchers in related studies. Finally, the conclusion and future research direction are presented in Section 6.

# 2 Literature Review

Feature selection has been widely used in many domains: text categorization [19], genomic analysis [20], [21], intrusion detection [22], [23], bioinformatics [24] among others. This chapter discussed some of the application of feature selection in intrusion detection. In this domain, many researchers applied feature selection seeking to come up with new or better models for efficient and effective detection of network intrusion with minimal false detection. The work of Thaseen and Kumar [13] evaluates the classification ability of six distinct tree-based classifiers on the NSL-KDD dataset. They used WEKA's CONS and CFS filters to select 15/41 features of the dataset, however, no normalization was performed on the data (possibly because it has little impact on the performance of tree-based algorithms [25]). To evaluate the performance of the models, WEKA's evaluation measures were used and the RandomTree model achieves the highest accuracy and a reduced false alarm rate. Sivatha Sindhu et al., [26] proposed a lightweight IDS for multi-class categorization using a wrapper-based genetic algorithm for feature selection and a hybrid of neural network and decision tree (neurotree) for actual classification. They used 16/41 features of NSL-KDD datasets and min-max method to normalize the selected attributes. WEKA's evaluation measures were used for performance evaluation, and compared to tree-based single classifiers their proposed methods achieved the highest detection rate of 98.38%. Ghaffari Gotorlar et al., [27] also proposed a harmony search-support vector machine (HS-SVM) method for intrusion detection on a KSL-KDD dataset. They used the harmony search to select 21/41 best features and the selected numeric features were normalized using min-max method whereas the nominal values were converted to numeric. LibSVM library was used for training the SVM model. Detection rate and test time were used to evaluate the model performance and the results show that the proposed HS-SVM method overcomes the SVM drawback of time-consuming during testing. Setiawan et al., [28] proposed an IDS model using combination of feature selection method, normalization, and Support Vector Machine. WEKA's modified rank-based information gain filter was used to select 17/41 NSL-KDD dataset features and the numerical features were log normalized. The model was evaluated using WEKA's evaluation measures and they achieved an overall accuracy of 99.8%. Khammassi and Krichen [29] applied three distinct decision tree-based algorithms on a genetic algorithm-logistic regression wrapper selector (GALR-DT) in building separate IDS models. The three decision tree algorithms are C4.5, Random Forest, and Naïve Bayes Tree. They used wrapper approach on a genetic algorithm as the search strategy and logistic regression as the learning algorithm to select the best subset of features on KDDcup99 and UNSW-NB15 datasets. 18/41 features were selected in KDDcup99 and 20/42 features were selected in UNSW-NB15 datasets by the GA-LR wrapper. Log-scaling and Min-max of the 0-1 range were applied to normalize the data. Dataset-wise performance of the models were compared on detection rate, accuracy, and false alert rate. Their results show that UNSW-NB15 provides the lowest FAR with 6.39% and a good classification accuracy compared to KDDcup99 and thus, they conclude that the UNSW-NB15 dataset is more complex than the KDD99 dataset. Khan et al., [30] proposed a novel two-stage deep learning (TSDL) model based on a stacked auto-encoder with a soft-max classifier for efficient network intrusion detection. The modeling process comprises of two decision stages and the built model is capable of learning and classifying useful feature representations in a semi-supervised mode. They evaluate the effectiveness of their methods on KDD99 and UNSW-NB15 datasets. DSAE feature selection is used to select 10 features in each dataset and the selected features are normalized using the min-max method. Most used IDS model evaluation metrics were used to assess the performance of their proposed model; they achieve a high recognition rate of up to 99.996% and 89.134% for the KDD99 and UNSW-NB15 datasets respectively.

The findings from our review reveal certain limitations hinder the effect of feature selection in each of the work. The limitations can be classified in three categories: preprocessing stage (transformation - use of inappropriate encoding method, feature selection – use of few methods or no comparison made), modeling stage (use of few algorithms thus, no room for comparison and result validation), and dataset issues (use of outdated dataset). This study overcome these limitations. Since most of the selected ML algorithms in this work consider all features during training simultaneously, One-hot encoding, an approach suited for such ML algorithms [31], is used. One of the most common normalization method [32], Min-max, is also used. Five of the widely used ML algorithms in IDS modeling [5] are selected to enable comparison, and a contemporary UNSW-NB15 dataset [33] is selected for this study. A total of 25 models are developed and evaluated. In what follows next, the feature selection concept is introduced along with the proposed feature selection approach.

## 3    Feature Selection

Selecting the most useful and relevant features in a large dataset is an important means of reducing computational complexity and increasing efficiency of models. Feature Selection (FS) is one of the successful preprocessing techniques for selecting an optimal relevant subset of features from original features for building models. The optimality of the feature subset is measured by an evaluation criterion [34]. Feature selection reduces the number of features by removing irrelevant, redundant, or noisy data, and brings immediate effects on the subsequently built model [35].

### 3.1    FS General Procedure

As shown in Figure 1, a typical feature selection process consists of the following four basic steps [34]:

- *Subset generation*: produces candidate feature subsets for evaluation based on search starting point, direction, and strategy.
- *Subset evaluation*: evaluates and compares candidate subset with the preceding best subset based on certain evaluation criteria. If the new subset happens to be better, it replaces the previous best subset. The evaluation criteria can be independent (used in filter methods) or dependent (used in wrapper methods).
- *Stopping criterion*: controls the stoppage of the feature selection process. The process of subset generation and evaluation is repeated until a given stopping criterion is satisfied.
- *Result validation*: validates the selected best subset using prior knowledge or different tests by synthetic and/or real-world data sets.

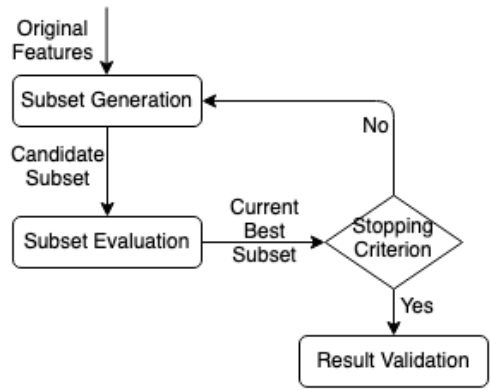

*Figure 1: Feature Selection Procedure*

## 3.2 Feature Selection Methods

Feature selection algorithms can be broadly categorized as a filter, wrapper, or embedded method [15], [34]. Several researchers have proposed the use of these methods in identifying important features for IDS [35].

- *Filter method*: relies on the general characteristics of the data to evaluate and select feature subsets. It separates feature selection from classifier learning so that the bias of a learning algorithm does not interact with the bias of a feature selection algorithm.
- *Wrapper method*: uses the predictive accuracy of a predetermined learning algorithm to determine the quality of selected features. Improves performance but in comparison to the filter method, it is computationally expensive to run for data with a large number of features.
- *Embedded method*: attempts to take advantage of the two methods by exploiting their different evaluation criteria in different search stages. It usually achieves comparable accuracy to the wrapper and comparable efficiency to the filter. It first incorporates the statistical criteria, as the filter method does, to select several candidate features subsets with a given cardinality, and then it chooses the subset with the highest classification accuracy as wrapper does. The embedded method performs both feature selection and model training simultaneously.

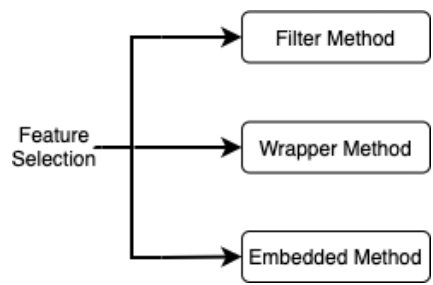

*Figure 2: Feature Selection Methods*

## 3.3 Proposed FS for IDS

Modern intrusion detection datasets inevitably contain plenty of redundant and irrelevant features which can affect the accuracy of IDS model [36], several researchers have tried to handle that using feature selection algorithms to select only the important intrusion features [35]. In this work, we propose the use of a wrapper-based feature selection approach with a decision tree algorithm as the feature evaluator to select optimal features and remove the inevitable plenty of redundant and irrelevant features in the dataset. Our proposal is based on the following reasons:

I. Most of the existing IDS datasets contain categorical features [8] and decision tree can handle both categorical and numeric features [25]
II. Decision tree is a low-bias algorithm [37], thus it can select optimal features while avoiding underfitting, which is one of the challenging issues in classification tasks [38]
III. Decision tree can be used to implement a trade-off between the performance of the selected features and the computation time which is required to find a subset [39]. Thus it can be stopped at any time, providing sub-optimal feature subsets.

## 3.4  FS Methods for Comparison

To assess the effectiveness of our proposed method, three filter selection methods are used for comparisons with the full-featured UNSW-NB15 dataset models as the baseline. A comparison with state-of-the-art results is also performed. Table 1 summarized the four feature selection methods. The explanation of our proposed method is given in the methodology section, a brief explanation of the three filter-based FS methods and their basic framework are provided below.

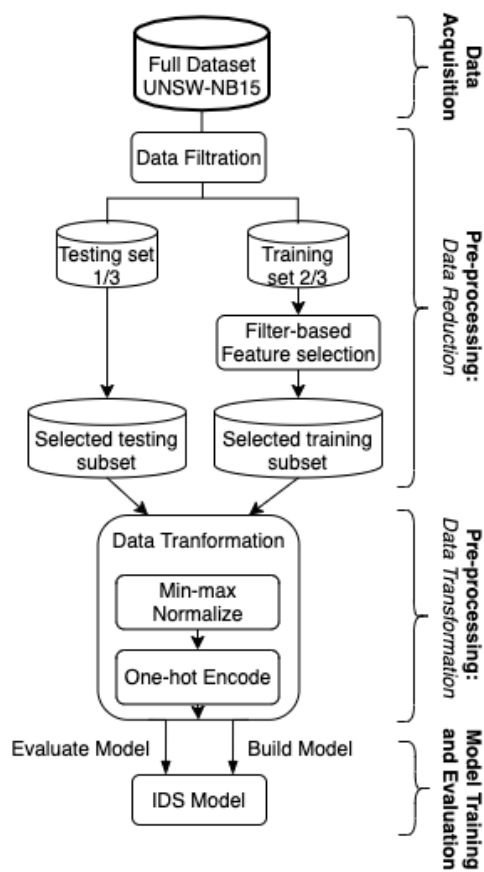

*Figure 3: Filter-based FS Framework and Model Training*

### 3.4.1  Information Gain (Info Gain or IG)

This is one of the most common feature evaluation techniques. IG evaluate the worth of a feature by measuring the expected reduction in information entropy with respect to the class [40]. The formula of the information gain is shown below:

$$\textit{Info Gain}\,(Class, Feature) = C(Class) - C(Class\,|\,Feature) \quad (1)$$

Where $C$ is the change in information entropy.

### 3.4.2 Gain Ratio (GR)

The Info Gain favors features with many values. The gain ratio seeks to avoid this bias by incorporating another term, split information, that is sensitive to how broadly and uniformly the considered data is split [40]. The gain ratio is defined as:

$$\textit{Gain Ratio}\,(Class, Feature) = \big(C(Class) - C(Class\,|\,Feature)\big) / C(Feature) \quad (2)$$

### 3.4.3 Relief Filter

This evaluates the worth of a feature by repeatedly sampling an instance and considering the value of the given feature for the nearest instance of the same and different classes. In other words, Relief estimates the quality of features according to how well their values distinguish between instances that are near each other. It can operate on both discrete and continuous class data [41].

*Table 1: Summary of the FS Methods*

| Name | FS Method | Feature Evaluator | Search Method |
|---|---|---|---|
| Decision Tree-Based | Wrapper | WrapperSubsetEval with J48 decision tree as a classifier | Bestfirst, forward |
| Information Gain | Filter | InfoGainAttributeEval | Ranker |
| Gain Ratio | Filter | GainRatioAttributeEval | Ranker |
| Relief Filter | Filter | ReliefAttributeEval | Ranker |

## 4 Methodology

This section described how the experiment is conducted by following the four basic machine learning steps (i.e. data acquisition, data preprocessing, model selection and training, and model evaluation). It also provides the tools used in the experiment.

### 4.1 Experimental Tools

In literature, many tools are used for implementing, evaluating, and comparing various IDS works. WEKA, general-purpose programming languages (such as Java, Python, etc.), and Matlab are the most used tools [5]. For this study, WEKA, Python in addition to Excel are used for data analysis and exploration, preprocessing, implementing, and validating the IDS models. Jupyter Notebook is used as the execution environment for Python and its libraries.

### 4.2 Dataset Acquisition

The UNSW-NB15 dataset which is among the latest and recommended dataset [8] and is found to be reliable, good for modern-day IDS modeling [17] is used in this work.

#### 4.2.1 UNSW-NB15 Dataset

The UNSW-NB15 dataset is a new IDS dataset created at the Australian Center for Cyber Security (ACCS) in 2015. About 2.5 million samples or 100GB of raw data were captured in modern network traffic including normal and attack behaviors and are simulated using the IXIA Perfect Storm tool and a tcpdump tool. 49 features were created using the Argus tool, the Bro-IDS tool, and 12 developed algorithms. The created features can be categorized into five groups: flow features, basic features, content features, time features, and additional generated features. The dataset has nine different modern attack types: Backdoor, DoS, Generic, Reconnaissance, Analysis, Fuzzers, Exploit, Shellcode, and Worms [18]. The UNSW-NB15 is considered as a new benchmark dataset that can be used for IDSs evaluation by the NIDS research community [42] and is recommended by [8]. For easy use and work reproducibility, the UNSW-NB15 comes along with predefined splits of a training set (175,341 samples) and a testing set (82,332 samples) [43], the predefined training and testing sets are used in this work. The publicly available training and testing set both contain only 44 features: 42 attributes and 2 classes. Since our primary focus is binary classification, the broad distribution of total attacks (anomaly) and normal traffic samples of the training and testing sets are used as shown in Table 2.

*Table 2: UNSW-NB15 Distribution Sample*

| Category | Training Set | | Testing Set | |
| --- | --- | --- | --- | --- |
| | Size | Distribution (%) | Size | Distribution (%) |
| Total Attacks | 119,341 | 68.06 | 45,332 | 55.06 |
| Normal | 56,000 | 31.94 | 37,000 | 44.94 |
| **Overall Samples** | **175,341** | **100** | **82,332** | **100** |

## 4.3 Data Preprocessing

In this study, two major preprocessing steps are used, namely, data reduction (filtration and feature selection) and data transforming (data normalization and encoding).

### 4.3.1 Data Reduction

#### 4.3.1.1 Data filtration

Irrelevant data is removed in both the training and testing set. The UNSW-NB15 dataset comes with 42 attributes, 2 class attributes, and an additional id attribute that is removed. Furthermore, since we are only interested in binary classification, the class attribute *attack_cat* indicating the categories of attacks and normal traffic is removed before feature selection.

#### 4.3.1.2 Feature Selection

The training set is used in feature selection to avoiding information leakage and subsequent building of misleading or overfitting models [44], the testing set is solely used for models' performance assessments.

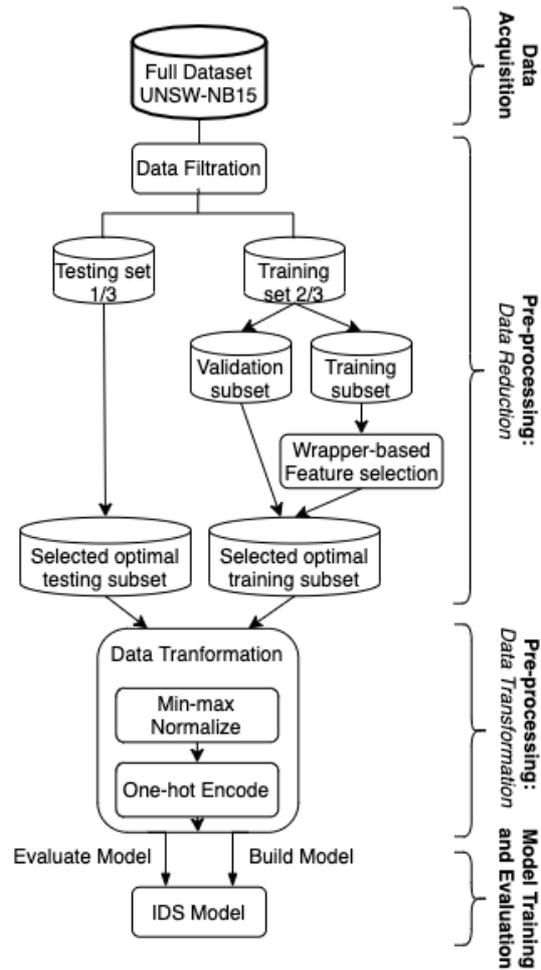

*Figure 4: Wrapper-based FS and Model Training Framework*

We propose the use of a wrapper-based DT approach where *BestFirst* Forward search strategy is used in feature search with 5 consecutive non-improving nodes as the search stopping criteria and accuracy as the evaluation measure. The J48, a java implementation of Quinlan's C4.5 [45], decision tree algorithm [44] provided in WEKA [46] is used as the feature evaluator. Our proposed approach selected 19 optimal features, and Figure 4 depicts the wrapper feature selection and the entire modeling process. For the filter methods, the default WEKA evaluator and *Ranker* search method setup of each filter is used. Since the filter methods rank all the features by their evaluations, the 19 top-ranked features are selected in each, and Figure 3 above depicts the filter feature selection and their modeling process. After the feature selection operations, WEKA's supervised attribute *Remove* filter is used to collect the features subsets in all the feature selection methods. Table 3 shows the selected features.

*Table 3: The Selected Features*

| FS Method | FS No. | Selected Features |
| --- | --- | --- |
| DT-Based Wrapper | 19 | *proto, *service, spkts, sbytes, dbytes, dttl, sloss, dloss, swin, stcpb, trans_depth, response_body_len, ct_srv_src, ct_src_dport_ltm, ct_dst_sport_ltm, ct_dst_src_ltm, ct_flw_http_mthd, ct_src_ltm, ct_srv_dst |
| Info Gain Filter | 19 | sbytes, dbytes, sttl, dttl, ct_state_ttl, rate, sload, smean, dur, dmean, dinpkt, dpkts, dload, sinpkt, tcprtt, synack, ackdat, sjit, spkts |
| Gain Ration Filter | 19 | sttl, dttl, ct_state_ttl, is_sm_ips_ports, *state, ackdat, tcprtt, synack, dinpkt, dload, dbytes, dpkts, rate, sbytes, dmean, dur, ct_dst_sport_ltm, response_body_len, smean |
| Relief Filter | 19 | *service, *proto, dttl, sttl, ct_dst_sport_ltm, smean, ct_state_ttl, ct_dst_ltm, ct_src_ltm, ct_src_dport_ltm, dload, ct_srv_dst, ct_srv_src, rate, ct_dst_src_ltm, dmean, is_sm_ips_ports, dtcpb, stcpb |
| (*) – indicates nominal features | | |

### 4.3.2 Data Transformation

#### 4.3.2.1 Data Normalization

The UNSW-NB15 dataset consists of two types: numeric and nominal features. To avoid classifier bias towards numeric features with large value ranges, min-max normalization with a range of 0 to 1 is applied on all the numeric features across the datasets using the Equation (3) Below. The normalization process is performed after feature selection in order not to affect the selection process.

$$x_{new} = \frac{x - min(x)}{max(x) - min(x)} \tag{3}$$

#### 4.3.2.2 Data Encoding

All the categorical (nominal) features are one-hot encoded. The full UNSW-NB15 dataset has 39 numeric and 3 nominal features, the nominal features are *proto*, *service*, and *state*. Two nominal features (*proto*, and *service*) are selected by the DT-based wrapper and the Relief filter, whereas the Gain Ration filter selects 18 numeric and only one nominal feature (state). All the 19 features selected by the Information Gain filter are numeric. An example of one-hot encoding of *protocol_type* feature with three sample values is shown in Table 4. Because one-hot encoding increases the dataset dimension, to avoid losing some nominal features' values encoded during feature selection, the encoding is performed after the feature selection and normalization processes.

*Table 4: One-Hot Encoding Example*

| Protocol_type |
|---|
| UDP |
| TCP |
| ICMP |

| UDP | TCP | ICMP |
|---|---|---|
| 1 | 0 | 0 |
| 0 | 1 | 0 |
| 0 | 0 | 1 |

After encoding the features, the dimensions of the final datasets increased as shown in Table 5. The final encoded features are then used in training the models.

*Table 5: Final Datasets Dimensions*

| Dataset | UNSW-NB15 Dataset features | |
|---|---|---|
| | Before Encoding | After Encoding |
| Full dataset | 42 | 194 |
| DT Wrapper | 19 | 163 |
| IG Filter | 19 | 19 |
| GR Filter | 19 | 27 |
| Relief Filter | 19 | 163 |

## 4.4 Model Selection and Training

Building the models constitutes of two stages: training stage and testing stage. So, the dataset is divided into two sets, the training set and testing set using the hold-out method. During the training stage, the algorithms are trained using the training set, then in the testing stage, the testing set is used to assess the performance and reliability of the built IDS models. Figure 3 and Figure 4 depicted the entire model training and testing processes. Using the full and the various FS datasets, the selected algorithms are used to build a total of 25 models. To measure the effectiveness of our FS methods, some evaluation metrics are used to evaluate and compare the models. The model evaluation metrics and the result of the evaluations are provided in the next subsection and Result and Discussion section of this work respectively.

## 4.5 Model Evaluation Metrics

Most of the IDS works made use of three metrics, namely; classification accuracy, detection rate (DR), and false alarm rate (FAR) [47]. Similarly, in this work, these metrics are adopted in addition to computational time. The formulas associated with the metrics are:

$$Accuracy(ACC) = \frac{TP + TN}{TP + TN + FP + FN} \quad (4)$$

$$Detection\ Rate(DR) = \frac{TP}{TP + FN} \quad (5)$$

$$False\ Alert\ Rate(FAR) = \frac{FP}{FP + TN} \quad (6)$$

The computational time is the entire time taken to train and evaluate a model, including the FS time. Because the timing depends on factors beyond our control such as CPU task switching, etc., we try to avoid running heavy tasks whilst executing the programs, we also try preventing the computer from sleeping to ensure minimal interference.

# 5 Result and Discussion

This section presents the platform on which the experiment is performed, the result obtained, and the interpretation of the results. We compare our proposed approach with other feature selection approaches and finally, we made comparisons with state-of-the-art IDS works.

## 5.1 Experimental Platform

To avoid interference from the experimental platform, all the models are implemented and executed in the same environment using the same programming language as shown in Table 6.

*Table 6: Experimental Platform*

| Name | Details |
|---|---|
| Computer | Lenovo ThinkPad T450 |
| OS | Windows 7 Ultimate 64-bit |
| CPU | 2.30GHz Intel Core i5 series 5 processor |
| RAM | 8GB (7.70GB usable) |
| Storage Disk | 240GB SSD |

| Execution platform | Jupyter Notebook |
|---|---|
| Experimental Tools | Excel, WEKA, Python |

## 5.2 Performance and Computational Time Comparisons

Five models are built with each of the selected algorithms. In this sub-section, comparisons of models built using our proposed method and those built using the three filter-based feature selection methods are made with the models built using the full features of the UNSW-NB15 dataset as the baseline models. The basis of the comparisons is the performance and the computation time shown in Table 7 and Table 8 respectively.

In ANN models, our methods perform rather poorly achieving third-best score on accuracy and FAR with third-highest computation time. It improves computation time but not the performance in comparison to the baseline model. In comparison to the baseline and the filter-based methods, our method achieves the worst performance on SVM across all metrics with the highest overall computational time of all models. Against the baseline model, our method improves neither performance nor computational time, making it the worst SVM model. Our method also performed poorly on the KNN model achieving the worst on accuracy and FAR as well as third-best on DR and computational time against all other KNN models. It thus, improve on computational time but not on performance scores. The proposed method achieved its best performance on the RF model with an accuracy of 86.41%, which is the third-best, after baseline and relief filter-based models. It achieves the best FAR with third-best computation time with two of the filter-based models taking less computation time. Our method failed to improve model performance against the baseline but it does improve on computational time. With NB models, our method achieves similar performance to the baseline model in lower time, and against other methods, though our method achieves joint best FAR on NB models, it has the second-lowest detection rate (DR), which is more important than the other metrics in IDS [48]. In terms of computational time, our method has the third-best, after IG and GR filter-based models which also achieves worst on FAR than our method.

*Table 7: Models Performance Comparisons*

| Models | Evaluation metrics | UNSW-NB15 | | | | |
|---|---|---|---|---|---|---|
| | | Full Features | DT Wrapper | IG Filter | GR Filter | Relief Filter |
| ANN | ACC | 86.00 | 82.08 | 82.06 | 83.72 | **86.51** |
| | DR | 98.62 | 97.94 | 99.41 | 98.39 | **97.99** |
| | FAR | 29.45 | 37.36 | 39.19 | 34.26 | **27.56** |
| SVM | ACC | **81.6** | 79.11 | 80.87 | 80.92 | 81.58 |
| | DR | **99.64** | 99.31 | 99.84 | 99.87 | 99.6 |
| | FAR | **40.51** | 45.64 | 42.38 | 42.29 | 40.5 |
| KNN | ACC | 84.78 | 83.21 | 86.1 | **86.96** | 84.81 |
| | DR | 96.46 | 96.44 | 96.01 | **95.9** | 96.52 |
| | FAR | 29.53 | 33.01 | 26.04 | **23.98** | 29.53 |
| RF | ACC | **86.82** | *86.41* | 86.14 | 85.98 | 86.49 |
| | DR | **98.7** | *97.95* | 97.8 | 98.00 | 98.79 |
| | FAR | **27.74** | *27.73* | 28.16 | 28.75 | 28.58 |
| NB | ACC | 55.61 | 55.61 | 76.37 | 71.28 | 55.44 |
| | DR | 19.39 | 19.38 | 93.7 | 98.16 | 19.07 |
| | FAR | 0.01 | 0.01 | 44.86 | 61.65 | 0.01 |

Overall, both the performance and computation time of models built using our method in comparison to the baseline and filter-based models are not satisfactory. Wrapper feature selection method is mainly known for its robustness in selecting best possible features to improve performance at the cost of a computation time [15], [34], with our proposed method, however, there is not any significant performance improvement despite the huge among of time taken in the feature selection process and in training the models which, on average, is higher than that of all the other methods. Thus, it can be deduced that, in the case of IDS modeling with UNSW-NB15 dataset, filter-based feature selection methods, like the ones used in this work, are better and should be considered as they performed fairly equivalent to our proposed wrapper method on corresponding models and also, they take lesser time as can be seen in Table 8 and Figure 5 below. Different wrapper feature evaluators, search approach, and/or termination condition can also be considered.

*Table 8: Models Computation Time Comparisons*

| Dataset Features | FS Time | Models Training Time | | | | | Overall Training Time Average |
|---|---|---|---|---|---|---|---|
| | | ANN | SVM | KNN | RF | NB | |
| Full Features | N/A | 11.27m | 181.68m | 17.92m | 0.74m | 4.64s | 42.34m |
| DT Wrapper | 31.4hrs | 4.95m | 259.1m | 10.94m | 0.63m | 2.86s | **55.13m** |
| IG Filter | 12.00s | 4.85m | 65.86m | 1.82m | 0.54m | 1.14s | 14.62m |
| GR Filter | 12.00s | 2.41m | 111.12m | 3.37m | 0.46m | 1.4s | 23.48m |
| Relief Filter | 3.92hrs | 8.00m | 193.65m | 16.79m | 0.65m | 3.08s | 43.83m |

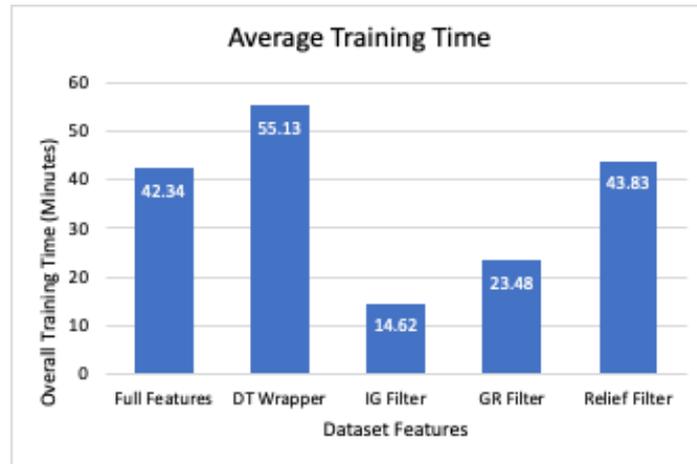

*Figure 5: Models' Average Training Time*

Furthermore, although the performance and computation time of some of the used algorithms can be influenced by other factors such as normalization [49], feature selection normally improves both performance and computation time of algorithms [15]. However, as seen in the comparisons made, these expectations failed to occur on many models with no significant improvement spotted on individual models against the baseline models. This essentially raises questions about the compliance nature of the UNSW-NB15 dataset.

## 5.3 Comparison with Other Works

To assess the effectiveness of our proposed method, we selected the best performing model, from among the models implemented using the method for corresponding comparisons with other state-of-the-art IDS works. There are many similar research works, we however limited our comparison to those that also used feature selection on the UNSW-NB15 dataset. We compare the percentages of accuracy (ACC), attack detection rate (DR), and false alert rate (FAR) whilst also paying attention to feature selection method, number of features, and algorithms used. Table 9 shows the performance comparisons of the works chronologically.

*Table 9: Related Works Comparisons*

| Work 'Year | FS Method | No. | Algorithm | ACC (%) | DR (%) | FAR (%) |
|---|---|---|---|---|---|---|
| [50] '15 | ARM-Based | 11 | LR | 83.0 | 68 | 14.2 |
| [29] '17 | GA-LR | 20 | DT | 81.42 | – | 6.39 |
| [51] '17 | Functional measures | 33 | DL-binomial | 98.99 | 95.84 | **0.56** |
| [30] '19 | DSAE | 10 | DL-Soft-max | 89.13 | – | 0.75 |
| [52] '19 | K-means | 41 | DNN | **99.19** | – | – |
| [53] '20 | NSGAII-ANN | 19 | RF | 94.8 | 94.8 | 6.0 |
| *This Work* | *DT-based* | *19* | *RF* | *86.41* | ***97.95*** | *27.73* |

From Table 9, it can be seen that our method achieves the best DR of 97.95%, performed better than two methods in ACC, and has the worst FAR. The best ACC and FAR are achieved by [52] and [51] respectively, both of which used deep learning classifiers. Their good results may be influenced by the use of deep learning classifiers which, recently, are proving to be good in IDS classification tasks [54]–[56]. It is important to note that in IDS, not detecting an attack can be costlier than mis-detecting an attack [48], thus DR can be more important than any other reported metrics and hence, our method can actually be more effective in detecting an attack than both [52] and [51]. Nonetheless, overall our method can be considered to be averagely good and fairly effective. Its major downside is the expensive computational time required, however, giving that IDSs are kind of systems that can be trained offline and deployed online for use [57], [58], this would not have been a major point of concern had our method improve performance against baseline models and also achieved better performance in comparison to filter-based methods.

# 6 Conclusion

Although our method is rated good in comparison to state-of-the-art works, we, however, discourage utilizing it in IDS modeling especially while working with UNSW-NB15 dataset giving that the filter methods used in this work achieved similar results at considerably lower feature selection and model training computation time; in addition to its failure to improve performance and reduce computational time as is generally expected with wrapper feature selection method. We recommend the use of different wrapper-based feature selection procedures or filter-based feature selection methods such as the ones used in this work in IDS modeling with the UNSW-NB15 dataset. Furthermore, our work highlighted the need for a more in-depth analysis of the conformity of the UNSW-NB15 dataset.

Some interesting and important future works can be done particularly on reducing the high false alert rate observed in our method because, besides a high detection rate, a good IDS should have a very low false alert rate. Furthermore, this work primarily focused on binary classification, however like most of the IDS datasets, the used dataset contained various attack types, thus multi-classification work can be done to further study the effectiveness of the proposed method.